\documentclass{JHEP3}

  \usepackage{latexsym,bm,amsmath,amssymb,amsfonts}
  \usepackage{epsfig,graphics,graphicx}
  \usepackage{slashed}
  \usepackage[latin1]{inputenc}
  \usepackage{mathrsfs}
\usepackage{mathcomp}
  
  \renewcommand{\theequation}{\thesection.\arabic{equation}}
  \long\def\comment#1{ }

 \newcommand{\bq}{\begin{equation}}
 \newcommand{\eq}{\end{equation}}
 \newcommand{\bqn}{\begin{eqnarray}}
 \newcommand{\eqn}{\end{eqnarray}}
 \newcommand{\nb}{\nonumber}
 \newcommand{\lb}{\label}

\title{\rm \LARGE \bf Lorentz Invariance Violation and Modified Hawking Fermions Tunneling from Black Strings}

\author{Shu-Zheng Yang$^{a}$, Kai Lin$^{b}$, Jin Li$^{c}$ and Qing-Quan Jiang$^{a}$\\

$^{a}$ Institute of Theoretical Physics, China
 West Normal University, Nanchong, Sichuan \\637002, People's Republic of
China\\
$^{b}$ Institute de F\'isica, Universidade de S\~ao
Paulo, CP 66318, 05315-970, S\~ao Paulo,
Brazil\\
$^{c}$
Department of Physics, Chongqing University,
Chongqing 400030, People's Republic of
China\\
{\tt E-mail address: szyangphys@126.com}}

\abstract{Recently the modified Dirac equation with Lorentz invariance
violation has been proposed, which would be helpful to resolve some
issues in quantum gravity theory and high energy physics. In this
paper, the modified Dirac equation has been generalized in curved
spacetime, and then fermion tunneling of black strings is researched
under this correctional Dirac field theory. We also use
semi-classical approximation method to get correctional
Hamilton-Jacobi equation, so that the correctional Hawking
temperature and correctional black hole's entropy are derived.}


\begin{document}

\section{Introduction}

In 1974, Hawking proved black holes could radiate Hawking radiation
once considering the quantum effect near the horizons of black holes
\cite{1}. This theory indicates black hole would be viewed as
thermodynamic system, so that black hole physics can be connected
closely with gravity, quantum theory and thermodynamics physics.
According to the view point of quantum tunneling theory, the virtual
particles inside of black hole could cross the horizon due to the
quantum tunneling effect, and become real particles, and then could
be observed by observers as Hawking radiation. Wilczek and Parikh et
al. proposed a semi-classical method to study the quantum tunneling
from the horizon of black hole \cite{2,3,4}. Along with this method,
the Hamilton-Jacobi method was applied to calculate the Hawking
tunneling radiation. According to the Hamilton-Jacobi method, the
wave function of Klein-Gordon equation can be rewritten as
$\Phi=C\exp\left(iS/\hbar\right)$ (where $S$ is semi-classical
action), and Hamilton-Jacobi equation is obtained via semi-classical
approximation. Using the Hamilton-Jacobi equation, the tunneling
rate could be calculated by the relationship
$\Gamma\sim\exp\left(-2\text{Im} S\right)$ (where $\Gamma$ is the
tunneling rate at the horizon of black hole), and then the Hawking
temperature can be determined. People have applied this method to
research Hawking tunneling radiation of several static, stationary
and dynamical black holes.

However, since the Hamilton-Jacobi equation is derived from
Klein-Gordon equation, original Hamilton-Jacobi method just can be
valid for scalar particles in principle. Therefore, Kerner and Mann
studied fermions tunneling of black hole by a new method \cite{5,6},
which assume the wave function of Dirac equation $\Psi$ as spin-up
and spin-down, and then calculate the fermions tunneling
respectively. Nevertheless, this method is still impossible to apply
in arbitrary dimensional spacetime. Our work in 2009 showed that,
the Hamilton-Jacobi equation can also be derived from Dirac equation
via semi-classical approximation, so that we proved the
Hamilton-Jacobi equation can be used to study the fermions tunneling
directly \cite{7}.

On the other hand, as the basis of general relativity and quantum
field theory, Lorentz invariance is proposed to be spontaneously
violated at higher energy scales. A possible deformed dispersion
relation is given by \cite{8}
 \bqn
 \lb{1}
p_0^2=\vec{p}^2+m^2-\left(Lp_0\right)^\alpha\vec{p}^2,
 \eqn
where $p_0$ and $\vec{p}$ are the energy and momentum of particle and
the $L$ is "minimal length" with the order of the Plank length. The
work of spacetime foam Liouville-string models have introduced this
relation with $\alpha=1$, and people also proposed quantum equation
of spinless particles by using this relation. Recently, Kruglov
considers the deformed dispersion relation with $\alpha=2$ and
proposes modified Dirac equation \cite{9}
 \bqn
 \lb{2}
\left[\bar{\gamma}^\mu\partial_\mu+m-iL\left(\bar{\gamma}^t\partial_t\right)\left(\bar{\gamma}^j\partial_j\right)\right]\psi=0,
 \eqn
where $\bar{\gamma}^a$ is ordinary gamma matrix, and $j$ is space
coordinate while $\mu$ is spacetime coordinate. The effect of the
correctional term would be observed in higher energy experiment.

In this paper, we try to generalize the modified Dirac equation in
curved spacetime, and then study the correction of Hawking tunneling
radiation. In section II, the modified Dirac equation in curved
spacetime is constructed and then the modified Hamilton-Jacobi
equation is derived via semi-classical approximation. We apply the
modified Hamilton-Jacobi equation to the fermions tunneling
radiation of $2+1$ dimensional black string and higher dimensional
BTZ-like black strings in section III and IV respectively, and the
section V includes some conclusion and the discussion about the
correction of black hole's entropy.

\section{Modified Dirac Equation and Hamilton-Jacobi Equation in Curved Spacetime} \label{sec1}

\renewcommand{\theequation}{2.\arabic{equation}} \setcounter{equation}{0}
As we all known, the gamma matrix and partial derivative should
become gamma matrix in curved spacetime $\gamma^a$ and covariant
${\cal D}_a$ derivative respectively, namely

 \bqn
 \lb{3}
\bar{\gamma}^a\rightarrow \gamma^a,~~~~~~\partial_a\rightarrow{\cal
D}_a=\partial_a+\Omega_a+\frac{i}{\hbar}eA_a,
 \eqn
where $\gamma^a$ satisfy the relationship
$\left\{\gamma^a,\gamma^b\right\}=\gamma^a\gamma^b+\gamma^b\gamma^a=2g^{ab}I$.
$eA_a$ is charged term of Dirac equation and
$\Omega_\mu=\frac{1}{8}\left(\gamma^a\gamma^b-\gamma^b\gamma^a\right)e^\nu_a\left(\partial_\mu
e_{b\nu}-\Gamma^c_{\mu\nu}e_{bc}\right)$ is spin connection.
According to this transformation, we can construct the modified
Dirac equation in curved spacetime as
 \bqn
 \lb{4}
\left[\gamma^\mu{\cal
D}_\mu+\frac{m}{\hbar}-\sigma\hbar\left(\gamma^t{\cal
D}^t\right)\left(\gamma^j{\cal D}^j\right)\right]\Psi=0,
 \eqn
where we choose $c=1$ but $\hbar\not=1$, while $c=\hbar=1$ in
Eq.(\ref{1}) and Eq.(\ref{2}). It is assumed that $\sigma\ll1$, so
that the correctional term $\sigma\hbar\left(\gamma^t{\cal
D}^t\right)\left(\gamma^j{\cal D}^j\right)$ is very small.

Now let's use the modified Dirac equation to derive the modified
Hamilton-Jacobi equation. Firstly, we rewrite the wave function of
Dirac equation as \cite{7}
 \bqn
 \lb{5}
\Psi=\zeta\left(t,x^j\right)\exp\left[\frac{i}{\hbar}S\left(t,x^j\right)\right],
 \eqn
where $\zeta\left(t,x^j\right)$ and $\Psi$ are $m\times1$ matrices,
and $\partial_tS=-\omega$. In semi-classical approximation, we can
consider the $\hbar$ is very small, so that we can neglect the terms
with $\hbar$ after dividing by the exponential terms and multiplying
by $\hbar$. Therefore, Eq.(\ref{4}) is rewritten as
 \bqn
 \lb{6}
\left[i\gamma^\mu\left(\partial_\mu
S+eA_\mu\right)+m-\sigma\gamma^t\left(\omega-eA_t\right)\gamma^j\left(\partial_jS+eA_j\right)\right]\zeta\left(t,x^j\right)=0.
 \eqn
 Considering the relationship
 \bqn
 \lb{7}
\gamma^\mu\left(\partial_\mu
S+eA_\mu\right)=-\gamma^t\left(\omega-eA_t\right)
+\gamma^j\left(\partial_jS+eA_j\right),
 \eqn
we can get
 \bqn
 \lb{8}
\left[i\Gamma^\mu\left(\partial_\mu
S+eA_\mu\right)+M\right]\zeta\left(t,x^j\right)=0,
 \eqn
where
 \bqn
 \lb{9}
\Gamma^\mu&=&\left[1+i\sigma\left(\omega-eA_t\right)\gamma^t\right]\gamma^\mu,\nb\\
M&=&m-\sigma g^{tt}\left(\omega-eA_t\right)^2.
 \eqn
Now, multiplying both sides of Eq.(\ref{9}) by the matrix
$-i\Gamma^\nu\left(\partial_\nu S+eA_\nu\right)$, so that we can
obtain
 \bqn
 \lb{10}
\Gamma^\nu\left(\partial_\nu
S+eA_\nu\right)\Gamma^\mu\left(\partial_\mu
S+eA_\mu\right)\zeta-iM\Gamma^\nu\left(\partial_\nu
S+eA_\nu\right)\zeta=0
 \eqn
The second term of above equation could be simplified again by
Eq.(\ref{8}), so above equation can be rewritten as
 \bqn
 \lb{11}
\Gamma^\nu\Gamma^\mu\left(\partial_\nu
S+eA_\nu\right)\left(\partial_\mu S+eA_\mu\right)\zeta+M^2\zeta=0,
 \eqn
where we can prove the relation
 \bqn
 \lb{12}
\Gamma^\nu\Gamma^\mu=\gamma^\nu\gamma^\mu+2i\sigma\left(\omega-eA_t\right)
g^{t\nu}\gamma^\mu+{\cal O}(\sigma^2).
 \eqn
We always ignore ${\cal O}(\sigma^2)$ terms because $\sigma$ is very
small. Now, let's exchange the position of $\mu$ and $\nu$ in
Eq.(\ref{11}), and consider the relation of gamma matrices
$\left\{\gamma^a,\gamma^b\right\}=2g^{ab}I$, then we can obtain
 \bqn
 \lb{13}
 \bigg\{\frac{\gamma^\alpha\gamma^\beta+\gamma^\alpha\gamma^\beta}{2}\left(\partial_\alpha
S+eA_\alpha\right)\left(\partial_\beta S+eA_\beta\right)+m^2-2\sigma
m
g^{tt}\left(\omega-eA_t\right)^2&&\nb\\
+2i\sigma\left(\omega-eA_t\right) g^{t\rho}\left(\partial_\rho
S+eA_\rho\right)\gamma^\mu\left(\partial_\mu
S+eA_\mu\right)\big\}\zeta+{\cal O}(\sigma^2)&&\nb\\
=\big\{g^{\alpha\beta}\left(\partial_\alpha
S+eA_\alpha\right)\left(\partial_\beta S+eA_\beta\right)+m^2-2\sigma
m
g^{tt}\left(\omega-eA_t\right)^2&&\nb\\
+2i\sigma\left(\omega-eA_t\right) g^{t\rho}\left(\partial_\rho
S+eA_\rho\right)\gamma^\mu\left(\partial_\mu
S+eA_\mu\right)\big\}\zeta+{\cal O}(\sigma^2)&=&0.
 \eqn
Namely
 \bqn
 \lb{13a}
\left[i\sigma\gamma^\mu\left(\partial_\mu S+eA_\mu\right)+{\cal
M}\right]\zeta\left(t,x^j\right)=0,
 \eqn
where
 \bqn
 \lb{13b}
{\cal M}=\frac{g^{\alpha\beta}\left(\partial_\alpha
S+eA_\alpha\right)\left(\partial_\beta S+eA_\beta\right)+m^2-2\sigma
m g^{tt}\left(\omega-eA_t\right)^2}{2 g^{t\rho}\left(\partial_\rho
S+eA_\rho\right)\left(\omega-eA_t\right)}.
 \eqn
Using the idea of Eq.(\ref{10})-(\ref{11}) again, we can multiply
both sides of Eq.(\ref{13b}) by the matrix
$-i\gamma^\nu\left(\partial_\nu S+eA_\nu\right)$, so that the
equation becomes
 \bqn
 \lb{13c}\sigma\gamma^\nu\left(\partial_\nu
S+eA_\nu\right)\gamma^\mu\left(\partial_\mu
S+eA_\mu\right)\zeta-i{\cal M}\gamma^\nu\left(\partial_\nu
S+eA_\nu\right)\zeta=0.
 \eqn
The second term of above equation could be simplified again by
Eq.(\ref{13b}). Then, exchange $\mu$ and $\nu$, and use the
relationship $\left\{\gamma^a,\gamma^b\right\}=2g^{ab}I$, so above
equation can be rewritten as
 \bqn
 \lb{14}
 \left[\frac{\gamma^\nu\gamma^\mu+\gamma^\mu\gamma^\nu}{2}
\sigma^2\left(\partial_\nu S+eA_\nu\right)\left(\partial_\mu
S+eA_\mu\right)+{\cal M}^2\right]\zeta\left(t,x^j\right)&&\nb\\
=\left[\sigma^2g^{\mu\nu}\left(\partial_\nu
S+eA_\nu\right)\left(\partial_\mu S+eA_\mu\right)+{\cal
M}^2\right]\zeta\left(t,x^j\right)&=&0.
 \eqn
The condition that Eq.(\ref{14}) has non-trivial solution required
the determinant of coefficient in Eq.(\ref{14}) should vanish, so we
can directly get the equation
 \bqn
 \lb{14a}
\sigma^2g^{\mu\nu}\left(\partial_\nu
S+eA_\nu\right)\left(\partial_\mu S+eA_\mu\right)+{\cal M}^2=0.
 \eqn
Consider the square root for left side of Eq.(\ref{14a}) and ignore
all the ${\cal O}\left(\sigma^2\right)$ terms, so we can directly
get the modified Hamilton-Jacobi equation:
 \bqn
 \lb{15}
g^{\mu\nu}\left(\partial_\nu S+eA_\nu\right)\left(\partial_\mu
S+eA_\mu\right)+m^2-2\sigma mg^{tt}\left(\omega-eA_t\right)^2=0.
 \eqn

 Therefore, we find the modified Dirac equation from Lorentz
invariance violation could lead to the modified Hamilton-Jacobi
equation, and the correction of Hamilton-Jacobi equation depends on
the energy and mass of radiation fermions. Using the modified
Hamilton-Jacobi equation, we then investigate the fermions Hawking
tunneling radiation of $2+1$ dimensional black string and $n+1$
dimensional BTZ-like string in following two sections.

\section{Fermions tunneling of 2+1 dimensional black string} \label{sec2}
\renewcommand{\theequation}{3.\arabic{equation}} \setcounter{equation}{0}

The research of gravity in 2+1 dimension can help people further understand the properties of gravity, and it is also
important to construct the quantum gravity. Recently, Murata, Soda
and Kanno have researched the $2+1$ dimensional gravity with dilaton
field, which action is given by \cite{10}
 \bqn
 \lb{16}
I=M_3\int d^3x\sqrt{-g}\left(BR+\frac{\lambda^2}{B}\right),
 \eqn
where, $B$, $M_3$ and $\lambda$ are respectively the dilaton field,
3-dimensional Planck mass and the parameter with mass dimension. The
static black string solution is given by
 \bqn
 \lb{17}
ds^2&=&-\ln\left(\frac{r}{r_H}\right)dt^2+\ln\left(\frac{r}{r_H}\right)^{-1}dr^2+dy^2,\nb\\
B&=&\lambda r
 \eqn
It is evident that the horizon of this black hole is $r_H$, but the
black string is unstable as $r_H\lesssim {\cal L}$ (where ${\cal L}$
is scale of compactification), so it is assumed that $r_H\gg {\cal
L}$.

Now we research the fermions tunneling of this black hole, so the
modified Hamilton-Jacobi equation in this spacetime is given by
 \bqn
 \lb{18}
-(1-2\sigma
m)\ln\left(\frac{r}{r_H}\right)^{-1}\omega^2+\ln\left(\frac{r}{r_H}\right)\left(\frac{dR}{dr}\right)^2+\left(\frac{dY}{dy}\right)^2+m^2=0,
 \eqn
where we have set $S=-\omega t+R(r)+Y(y)$, yielding the radial
Hamilton-Jacobi equation as
 \bqn
 \lb{19}
-(1-2\sigma
m)\ln\left(\frac{r}{r_H}\right)^{-1}\omega^2+\ln\left(\frac{r}{r_H}\right)\left(\frac{dR}{dr}\right)^2+\lambda_0+m^2=0,
 \eqn
where the constant $\lambda_0$ in above equation is from separation
of variables, and Eq.(\ref{19}) finally can be written as
 \bqn
 \lb{20}
R_\pm(r)=\pm\int\ln\left(\frac{r}{r_H}\right)^{-1}\sqrt{(1-2\sigma
m)\omega^2-\ln\left(\frac{r}{r_H}\right)\left(\lambda_0+m^2\right)},
 \eqn
At the horizon $r_H$ of the black string, above equation is
integrated via residue theorem, and we can get
 \bqn
 \lb{21}
R_\pm(r)=\pm i\pi\left(1-\sigma m\right)r_H\omega,
 \eqn
and the fermions tunneling rate
 \bqn
 \lb{22}
\Gamma&=&\exp\left(-\frac{2}{\hbar}\text{Im}S\right)=\exp\left[-\frac{2}{\hbar}\left(\text{Im}R_+-\text{Im}R_-\right)\right]\nb\\
&=&e^{-\frac{4\pi}{\hbar}(1-\sigma
m)r_H\omega}=e^{-\frac{\omega}{T_H}}.
 \eqn
The relationship between tunneling rate and Hawking temperature
required
 \bqn
 \lb{23}
T_H=\hbar\frac{1+\sigma m}{4\pi r_H}=(1+\sigma m)T_0.
 \eqn
where $T_H$ and $T_0$ are the modified and non-modified Hawking
temperature of the $2+1$ dimensional black string respectively, and
$\sigma$ term is the correction.

\section{Fermions tunneling of higher dimensional BTZ-like black strings } \label{sec3}
\renewcommand{\theequation}{4.\arabic{equation}} \setcounter{equation}{0}

As we all know that the linear Maxwell action fails to satisfy the
conformal symmetry in higher-dimensional spacetime \cite{11}, so
Hassaine and Martinez proposed gravity theory with non-linear
Maxwell field in arbitrary dimensional spacetime
 \bqn
 \lb{24}
I=-\frac{1}{16\pi}\int_Md^{n+1}x\sqrt{-g}\left[R+\frac{2}{l^2}-\beta\left(\alpha
F_{\mu\nu}F^{\mu \nu}\right)^s\right].
 \eqn
where $\Lambda\equiv-l^{-2}$ is cosmological constant. Hendi
researched $n+1$ dimensional static black strings solution with
$\beta=1$, $\alpha=-1$ and $s=n/2$, and it is charged BTZ-like
solutions \cite{12}, which metric is
 \bqn
 \lb{25}
ds^2=-f(r)dt^2+\frac{dr^2}{f(r)}+r^2\sum\limits_{k}\left(dx^k\right)^2,
 \eqn
where
 \bqn
 \lb{26}
f(r)=\frac{r^2}{l^2}-r^{2-n}\left(M+2^{n/2}Q^{n-1}A_t\right),
 \eqn
and the electromagnetic potential
 \bqn
 \lb{27}
A=A_tdt=Q\ln\left(\frac{r}{l}\right)dt
 \eqn
As $n=2$, this solution is no other than the static charged BTZ
solution. We will study the Hawking radiation and black hole
temperature at the event horizon $r_H$ of this black string. In
Eq.(\ref{15}), we can set $S=-\omega t+R(r)+Y(x^k)$, where $x^k$ are
the space coordinates excluding the radial coordinate, so that the
modified Hamilton-Jacobi equation is given by
 \bqn
 \lb{28}
-(1-2\sigma
m)f^{-1}(r)\left(\omega-eA_t\right)^2+f(r)\left(\frac{dR}{dr}\right)^2+\frac{1}{r^{2}}\sum\limits_k\left(\frac{dY}{dx^k}\right)^2+m^2=0,
 \eqn
and the radial equation with constant $\lambda_0$ is
 \bqn
 \lb{29}
-(1-2\sigma
m)f^{-1}(r)\left(\omega-eA_t\right)^2+f(r)\left(\frac{dR}{dr}\right)^2+\frac{\lambda_0}{r^{2}}+m^2=0,
 \eqn
Therefore, at the horizon $r_H$ of the black string, $f(r_H)=0$ and
we finally get
 \bqn
 \lb{30}
R_\pm(r)&=&\pm\int f(r)^{-1}\sqrt{(1-2\sigma
m)\left(\omega-eA_t\right)^2-f(r)\left(\lambda_0+m^2\right)}\nb\\
&=&\pm
i\pi(1-\sigma m)\frac{\omega-\omega_0}{f'(r_H)},
 \eqn
where $\omega_0=eA_t(r_H)$. It means that the fermions tunneling
rate is
 \bqn
 \lb{31}
\Gamma&=&\exp\left(-\frac{2}{\hbar}\text{Im}S\right)=\exp\left[-\frac{2}{\hbar}\left(\text{Im}R_+-\text{Im}R_-\right)\right]\nb\\
&=&e^{-\frac{4\pi}{\hbar}(1-\sigma
m)\frac{\omega-\omega_0}{f'(r_H)}}=e^{-\frac{\omega-\omega_0}{T_H}}.
 \eqn
and the Hawking temperature is
 \bqn
 \lb{32}
T_H=\hbar\frac{1+\sigma m}{4\pi }f'(r_H)=\hbar\frac{1+\sigma
m}{4\pi}\left(\frac{nr_H}{l^2}-2^{\frac{n}{2}}Q^nr_H^{1-n}\right)=(1+\sigma
m)T_0.
 \eqn
where $T_H$ and $T_0$ are the modified and non-modified Hawking
temperature of the $n+1$ dimensional BTZ-like black string
respectively , and $\sigma$ term is the correction.

\section{Conclusions}
\renewcommand{\theequation}{5.\arabic{equation}} \setcounter{equation}{0}

In this paper, we consider the deformed dispersion relation with
Lorentz invariance violation, and generalize the modified Dirac
equation in curved spacetime. The fermions tunneling radiation of
black strings is researched, and we find the modified Dirac equation
could lead to the Hawking temperature's correction, which depend on
the correction parameter $\sigma$ and particle mass $m$ in the
modified Dirac equation. Next we will discuss the correction of
black hole entropy in this theory.

The first law of black hole thermodynamics require
 \bqn
 \lb{33}
dM=TdS+\Xi dJ+UdQ,
 \eqn
where $\Xi$ and $U$ are electromagnetic potential and rotating
potential, so the non-modified entropy of black hole is \cite{1, 13}
 \bqn
 \lb{34}
dS_0=\frac{dM-\Xi dJ-UdQ}{T_0}.
 \eqn
From above results, we know the relationship between modified and
non-modified Hawking temperature is $T_H=(1+\sigma m)T_0$, since the
non-modified black hole entropy is given by
 \bqn
 \lb{35}
S_H=\int dS_H=\int \frac{dM-\Xi dJ-UdQ}{(1+\sigma m)T_0}=S_0-m\int
\sigma dS_0+{\cal O}(\sigma^2),
 \eqn
where we can ignore the ${\cal O}(\sigma^2)$ because $\sigma\ll 1$.
Eq.(\ref{35}) shows that the correction of black hole entropy
depends on $\sigma$, which is independent from time and space
coordinates. However, it is possible that $\sigma$ depends on other
parameters in curved spacetime, and it is very interesting that
$\sigma$ depends on $S_0$. Especially, as
$\sigma=\frac{\sigma_0}{S_0}+\cdot\cdot\cdot$, we can get the
logarithmic correction of black hole entropy
 \bqn
 \lb{36}
S_H=S_0-m\sigma_0 \ln S_0+\cdot\cdot\cdot.
 \eqn
In quantum gravity theory, the logarithmic correction has been
researched in detail \cite{14,15}, and according to Ref.\cite{16},
it is required that the coefficient of logarithmic correction should
be $-\frac{n+1}{2(n-1)}$ in $n+1$ dimensional spacetime, so it
indicates the $\sigma_0$ could be $\frac{n+1}{2m(n-1)}$.

On the other hand, from the deformed dispersion relation (\ref{1})
with $\alpha=2$, it implies that the Klein-Gordon equation could be
given by
 \bqn
 \lb{37}
\left(-\partial^2_t+\partial^2_j+m^2-\sigma^2\hbar^2\partial^2_t\partial^2_j\right)\Phi=0,
 \eqn
so the generalized uncharged Klein-Gordon equation in static curved
spacetime is
 \bqn
 \lb{38}
\left[g^{tt}\nabla^2_t+g^{jj}\nabla^2_j+m^2+\sigma^2\hbar^2\left(g^{tt}\nabla^2_t\right)\left(g^{jj}\nabla^2_j\right)\right]\Phi=0,
 \eqn
and using the semi-classical approximation with
$\Phi=C\exp\left(iS/\hbar\right)$, the modified Hamilton-Jacobi
equation in scalar field is given by
 \bqn
 \lb{39}
\left(1+\sigma^2g^{tt}\omega^2\right)g^{\mu\nu}\partial_\mu
S\partial_\nu S+m^2-\sigma^2\left(g^{tt}\right)^2\omega^4=0,
 \eqn
namely
 \bqn
 \lb{40}
g^{\mu\nu}\partial_\mu S\partial_\nu
S+m^2-\sigma^2g^{tt}\omega^2\left(m^2+g^{tt}\omega^2\right)+{\cal
O}(\sigma^4)=0.
 \eqn
Contrasting Eq.(\ref{15}) and Eq.(\ref{40}) as uncharged case, we
find the correctional terms of Dirac field and scalar field are very
different. The fact implies that the corrections of Hawking
temperature and black hole entropy from Hawking tunneling radiation
with different spin particles could be different, and this
conclusion could be helpful to suggest a new idea to research the
black hole information paradox. Work in these fields is currently in
progress.

\section*{\bf Acknowledgements}
This work was supported by FAPESP No. 2012/08934-0, National Natural
Science Foundation of China No. 11205254, No. 11178018 and No.
11375279.


\end{document}